\documentclass[12pt,preprint]{aastex}

\begin{document}

\title{Who Pulled the Trigger: a Supernova or an AGB Star?}

\author{Alan P.~Boss and  Sandra A. Keiser}
\affil{Department of Terrestrial Magnetism, Carnegie Institution of
Washington, 5241 Broad Branch Road, NW, Washington, DC 20015-1305}
\authoremail{boss@dtm.ciw.edu, keiser@dtm.ciw.edu}

\begin{abstract}

 The short-lived radioisotope $^{60}$Fe requires production in a core 
collapse supernova or AGB star immediately before its incorporation into
the earliest solar system solids. Shock waves from a somewhat distant
supernova, or a relatively nearby AGB star, have the right speeds
to simultaneously trigger the collapse of a dense molecular cloud
core and to inject shock wave material into
the resulting protostar. A new set of FLASH2.5 adaptive mesh 
refinement hydrodynamical models shows that the injection efficiency 
depends sensitively on the assumed shock thickness and density. 
Supernova shock waves appear to be thin enough to inject the amount of 
shock wave material necessary to match the short-lived radioisotope 
abundances measured for primitive meteorites. Planetary nebula shock 
waves from AGB stars, however, appear to be too thick to achieve the 
required injection efficiencies. These models imply that a supernova 
pulled the trigger that led to the formation of our solar system.

\end{abstract}


\section{Introduction}

 Primitive meteorites contain daughter products of the decay of
short-lived radioisotopes (SLRIs) such as $^{26}$Al, $^{41}$Ca, $^{53}$Mn, 
and $^{60}$Fe, distributed in different minerals in a way 
that indicates the parent isotopes were still alive
at the time of their incorporation into the refractory inclusions
and chondrules that record the earliest history of the 
solar system. The presence of $^{60}$Fe is particularly significant,
as its production requires stellar nucleosynthesis (Tachibana \&
Huss 2003; Tachibana et al. 2006). Given half-lives on the order 
of $\sim 10^6$ yr, the evidence for these radioisotopes suggests
that the same stellar source that synthesized them may well have 
triggered the collapse of the presolar dense cloud core as well, 
while simultaneously injecting the freshly-synthesized radioisotopes
(Cameron \& Truran 1977; Boss 1995). Supernovae resulting from
massive stars in the range of $\sim 20 M_\odot$ to $\sim 60 M_\odot$ 
or planetary nebulae derived from intermediate-mass ($\sim 5 M_\odot$)
AGB stars have been proposed as possible sources of all or most of 
these radioisotopes (e.g., Huss et al. 2009).

 Shock-triggered collapse and injection into the presolar cloud 
(Cameron \& Truran 1977) has been proposed and studied in 
detail (e.g., Boss 1995; Foster \& Boss 1997; Vanhala \& Boss 2002; 
Boss et al. 2008, 2010). Recent calculations have shown that
simultaneous triggered gravitational collapse and injection
of shock wave gas and dust into the collapsing cloud core is possible
even when detailed heating and cooling processes in the shock-cloud 
interaction are included (Boss et al. 2008). Shock speeds in the 
range from 5 km/sec to 70 km/sec are capable of achieving simultaneous 
triggering and injection for a 2.2 $M_\odot$ target cloud 
(Boss et al. 2010). However, these models led to considerably lower 
injection efficiencies than those previously estimated on the basis 
of models where the shock-cloud interaction was assumed to be 
isothermal (Boss 1995; Foster \& Boss 1997; Vanhala \& Boss 2002).
When the injection efficiency ($f_i$) is defined to be the fraction 
of the incident shock wave material that is injected into the collapsing 
cloud core, values of $f_i \sim 0.001$ result from the nonisothermal
models (Boss et al. 2008, 2010), about 100 times lower than the
values of $f_i$ found previously for strictly isothermal 
interactions. Considering that the shock fronts in these models
contain 0.015 $M_\odot$ of gas and dust, this means that the 
Boss et al. (2008, 2010) models produced nominal dilution factors 
$D \sim 10^{-5}$, where $D$ is defined as the ratio of the amount 
of mass derived from the stellar source of the shock front that 
ends up in the protoplanetary disk to the amount of mass in the disk that 
did not derive from the stellar source. Such values appear to be much 
too low to explain the initial abundances inferred for typical SLRIs,
which range from $\sim 10^{-4}$ to $\sim 3 \times 10^{-3}$ for supernovae
(Takigawa et al. 2008; Gaidos et al. 2009) and $\sim 3 \times 10^{-3}$
for an AGB star (Trigo-Rodr\'iguez et al. 2009).

 Boss et al. (2010) found that varying the shock speed from
5 to 70 km/sec had relatively little effect on $f_i$, while doubling 
the density of the target cloud could decrease $f_i$ by a factor of 3.
Here we explore the effects of changes in the assumed shock
wave parameters, in order to learn if higher values of $f_i$ and
therefore $D$ might thereby result. In addition, we seek to learn
if these shock wave variations will indicate a preference for either 
a supernova or an AGB star wind for triggering the formation of 
the solar system.

\section{Numerical Methods}

 We used the FLASH2.5 code, as in our previous work (Boss et al. 2008, 
2010). FLASH2.5 advects gas using the piecewise parabolic method, 
accurate to second-order in space and time, with a Riemann solver 
at cell boundaries designed to handle strong shock fronts. Our tests 
of the FLASH2.5 code and further details about our implementation 
scheme are detailed in Boss et al. (2010). Basically, we used the 
two dimensional, cylindrical coordinate ($R, Z$) version of FLASH2.5, 
with axisymmetry about the rotational axis ($\hat z$). Multipole 
self-gravity was used, including Legendre polynomials up to $l = 10$. 
The cylindrical grid was typically 0.2 pc long in $Z$ and 0.063 pc
wide in $R$, though in some models the grid was extended to be 0.4 pc 
long in order to follow the evolution farther downstream. 
The number of blocks in $R$ ($N_{BR}$) was 5 in all cases, while 
the number of blocks in $Z$ ($N_{BZ}$) was 15 for the standard-length 
grids and 20 for the extended grids, with each block consisting 
of $8 \times 8$ grid points. The number of levels of grid refinement 
($N_L$) was 5 for all models.

 As in Boss et al. (2008, 2010), we included compressional heating
and radiative cooling, based on the results of Neufeld \& Kaufman (1993)
for cooling caused by rotational and vibrational transitions of 
optically thin, warm molecular gas composed of H$_2$O, CO, 
and H$_2$. As before, we assumed a radiative cooling rate of
$\Lambda \approx 9 \times 10^{19} (T/100) \rho^2$ erg cm$^{-3}$ s$^{-1}$,
where $T$ is the gas temperature in K and $\rho$ is the gas density 
in g cm$^{-3}$. The gas temperatures were constrained to lie in the range
between 10 K and 1000 K, as in Boss et al. (2008, 2010), based on
the results of Kaufman \& Neufeld (1996) for magnetic shock speeds 
in the desired range of 5 km/sec to 45 km/sec.

\section{Initial Conditions}

 The target dense cloud cores are modeled on Bonner-Ebert (BE) spheres
(Bonnor 1956), which are the equilibrium structures for self-gravitating,
isothermal spheres of gas. As in Boss et al. (2010), the BE-like
spheres are initially isothermal at 10 K, with a central density of
$1.24 \times 10^{-18}$ g cm$^{-3}$, a radius of 0.058 pc, a mass 
of 2.2 $M_\odot$, and are stable against collapse for at least $10^6$ yr.
The spheres are embedded in an intercloud medium with a density of 
$3.6 \times 10^{-22}$ g cm$^{-3}$ and a temperature of 10 K. 
Shock waves are launched downward from the top of the grid toward the
spheres (Figure 1) at a speed of 40 km/sec.
The standard shock front, as used in Boss et al.
(2008, 2010), has a thickness of 0.003 pc with a  uniform density 
of $3.6 \times 10^{-20}$ g cm$^{-3}$, a mass of 0.015 $M_\odot$, 
a temperature of 1000 K, and  is followed by a post-shock wind with a 
density of $3.6 \times 10^{-22}$ g cm$^{-3}$ and temperature of 1000 K, 
also moving downward at the same speed as the shock wave.
The shock front material is represented by a color field, initially 
defined to be equal to 1 inside the shock front and 0 elsewhere,
which allows the shock wave material to be tracked forward in time 
(e.g., Foster \& Boss 1997).

\section{Results}

 Table 1 lists the variations in the shock front parameters that
were explored in the new models as well as the resulting injection
efficiencies $f_i$ and dilution factors $D$. The models are all identical
except for the assumed properties of the initial shock front, where the
standard shock front densities of Boss et al. (2010) were multiplied
by factors ranging from 0.1 to 800, and the shock thickness by
factors ranging from 0.1 to 10. 

 Figure 1 shows the initial conditions for model 200-0.1, where
the initial shock density was 200 times the standard value and
the initial shock thickness was 0.1 times the standard value.
Figure 2 shows the shock-cloud interaction 0.04 Myr after Figure 1,
when the shock front has begun to drive the target cloud into
collapse: the maximum density has increased by nearly a factor
of 1000. Rayleigh-Taylor fingers have injected shock-front material
throughout most of the target cloud (as shown by the black contours
outlining the color field, initially in the shock front), while
Kelvin-Helmholtz instabilities are ablating the outer regions of
the cloud and transporting it downstream.  

 Figure 3 shows the 1000-AU-scale region around the dense, 
collapsing protostar whose collapse has been triggered by the shock 
front. The velocity vectors show that while more gas will be
accreted by the protostar, other gas is likely to be blown downstream
by the combination of the shock front and the post-shock wind and
will not be accreted. The mass of the protostar at this time
is $\sim 1 M_\odot$, implying that roughly half of the target cloud's
initial mass will be lost and half accreted by the protostar.
Figure 4 depicts the color field at the same time and on the same
spatial scale as Figure 3, showing that shock front material
has already been injected into the collapsing protostar, and
that more shock front material will be accreted as the collapse
proceeds. The bulk of the shock front material is swept downstream,
however. 

 The injection efficiency estimated for model 200-0.1 at the
time shown in Figures 3 and 4 is $f_i \approx 0.02$, while the
dilution factor for this model is $D \approx 3 \times 10^{-3}$.
Most of the models shown in Table 1 behaved in much the
same way as model 200-0.1, with the exception of the models
marked by asterisks. In these models, the shock front was so
vigorous that while the target cloud was compressed somewhat, 
the cloud did not reach a high enough density for dynamic,
self-gravitational collapse to begin, and by the time that the
cloud was pushed off the bottom of the numerical grid, the
shock had shredded the cloud more than it had triggered 
collapse. Hence, these models must be considered as failed
models, in spite of their high values of $f_i$ and $D$: evidently
the threat of shredding limits the injection efficiency.
Table 1 shows the important trends that for a fixed shock
density, increasing the shock thickness results in higher
dilution factors $D$, as does increasing the shock density
at fixed shock thickness, as might be expected, with cloud
shredding placing the ultimate limits on these trends.

 Given that $f_i \approx 0.02$ and $D \approx 3 \times 10^{-3}$
for model 200-0.1, values that are factors of $\sim 20$ and
$\sim$ 100 times higher than in the standard shock front model
(Boss 2010), respectively, it is clear that injection
efficiencies and dilution factors depend sensitively on
the assumed shock wave parameters, all other things being equal.

\section{Discussion}

 We now turn to the question of whether any of the injection efficiencies
and dilution factors shown in Table 1 are able to 
match the demands of the meteoritical record for the SLRIs, and
in particular, whether any such desirable shock waves might exist 
in reality.

\subsection{Supernova}

 The desired dilution factors for a supernova trigger range from
$D = 1.3 \times 10^{-4}$ to $1.9 \times 10^{-3}$ (Takigawa et al. 2008) 
to $D = 3 \times 10^{-3}$ (Gaidos et al. 2009). Table 1 shows
that four collapse models had $D$ values in this broad range:
models 100-0.1, 200-0.1, 400-0.1, and 10-1. However, these are not
the appropriate $D$ values for comparison with a supernova source,
because a supernova shock launched at $\sim$ 1000 km/sec must snowplow 
$\sim$ 25 times its own mass in order to slow down to $\sim$ 40 km/sec
(Boss et al. 2010). The model dilution factors in Table 1 must then be 
decreased by this same factor, dropping $D$ to $\sim 1.2 \times 10^{-4}$ 
for model 200-0.1 and $\sim 4 \times 10^{-4}$ for model 400-0.1. 
These values are close to those proposed by Takigawa et al. (2008),
but about 10 times smaller than that favored by Gaidos et al. (2009).
As noted by Boss et al. (2010), other factors can result in higher
values of $D$ for the models, such as incomplete accretion of
the target cloud (e.g., Figure 4, which would raise $D$ for model 200-0.1
by a factor of 2), preferential addition of the late arriving SLRIs
to the solar nebula, rather than the protosun, and lower target cloud
densities (and consequently larger initial cloud diameters). Given
that all of these factors work in the direction of increasing $D$,
the fact that both models 200-0.1 and 400-0.1 produce $D$ estimates much
closer to the desired range than the standard shock models (Boss et al.
2010) must be viewed as a positive outcome for a supernova trigger.

 However, a successful outcome demands that supernova shock waves in 
their radiative phase have properties similar to those of the shocks
assumed in models 200-0.1 and 400-0.1, where the shock thickness
was $10^{15}$ cm and the shock number densities were $2 \times 10^6$ 
cm$^{-3}$ and $4 \times 10^6$ cm$^{-3}$, respectively. The Cygnus Loop, 
the $\sim 10^4$-yr-old remnant of a core collapse (Type II) supernova, has a 
shock speed of 170 km/sec and a thickness no greater than $\sim 10^{15}$ cm 
(Blair et al. 1999), consistent with models 200-0.1 and 400-0.1. 
W44 is a $\sim 2 \times 10^4$-yr-old remnant of a Type II supernova,
where the shock fronts are colliding with giant molecular cloud (GMC) gas
with a density greater than $10^3$ cm$^{-3}$ (Reach, Rho, \& Jarrett 2005).
The shock front has slowed down to 20-70 km/sec and has thickened as a 
result of the GMC interaction, but is no thicker than $\sim 10^{17}$ cm.
For a nearly isothermal shock, the post-shock density $n_s$ for propagation
in a stationary medium of density $n_m$ is $n_s/n_m = (v_s/c_m)^2$, where 
$v_s$ is the shock speed and $c_m$ is the sound speed in the medium
(e.g., Spitzer 1968). For the present models, $v_s = 40$ km/sec,
$c_m = 0.2$ km/sec, and $n_m = 10^2$ cm$^{-3}$, leading to 
$n_s = 4 \times 10^6$ cm$^{-3}$. This is the same shock density
as used in model 400-0.1. Evidently, then, models 200-0.1 and 400-0.1 do 
appear to be reasonable models of evolved Type II supernova remnants similar
to the Cygnus Loop and W44, which have expanded to sizes of 10 pc
or more after $\sim 10^4$ yr of evolution.

\subsection{AGB star wind}

 Trigo-Rodr\'iguez et al. (2009) suggest that $D \sim 3 \times 10^{-3}$
is required for an AGB star source of the SLRIs. Only models
200-0.1 and 400-0.1 produced $D$ values at least this large. Note that
dilution caused by snowplowing does not need to be invoked here 
because planetary nebulae speeds are already in the proper range of
20 to 30 km/sec. However, the thickness of the planetary nebula
Abell 39 is estimated to be $\sim 3 \times 10^{17}$ cm 
(Jacoby, Ferland, \& Korista 2001) and for planetary nebula PFP-1 to be
$\sim 5 \times 10^{17}$ cm (Pierce et al. 2004). These thicknesses
are even greater than those in the models with 10 times the standard
shock thickness (Table 1), and so are incapable of producing
the desired dilution factor. Planetary nebulae appear to be
too thick to achieve the injection efficiencies needed to 
explain the solar system's SLRIs.

\subsection{Grain injection}

 The $D$ values in Table 1 are based on injection purely in
the gas phase, i.e., assuming that the SLRIs are either in the
gas phase or are locked up in grains small enough to remain
tied to the gas. As noted by Foster \& Boss (1997), large dust
grains can shoot through the gas of a stalled shock front as a result
of their momentum, thereby increasing the SLRI injection efficiency,
as studied by Ouellette et al. (2010). Hence the $D$ values in
Table 1 should be considered as lower bounds.

 The penetration distance of a dust grain with a radius $a_d$ 
and density $\rho_d$ moving in a gas of density $\rho_g$
can be estimated by the distance it must travel to impact an 
amount of gas equal to its own mass, thereby halving its speed.
This distance is $d = 4/3 (\rho_d/\rho_g) a$. The region at the
top of Figure 4 shows that dust grains might be preferentially injected 
if they could penetrate a distance of $d \sim 10^{16}$ cm into gas 
with a density $\rho_g \sim 10^{-18}$ g cm$^{-3}$. With 
$\rho_d = 2.5$ g cm$^{-3}$, this requires grains with a size
$a \sim 30$ $\mu$m or larger.
 
 The predicted power-law size distribution for dust grains 
formed by core collapse supernovae (e.g., Nozawa et al. 2003)
places most of the mass of the grains in the size range 
from 0.1 $\mu$m to 0.3 $\mu$m (Nath, Laskar, \& Shull 2008).
Presolar SiC grains of type "X" that originate in supernovae 
have sizes that fall in the range of 0.4 $\mu$m to 2 $\mu$m 
(Amari et al. 1994; Ouellette et al. 2010), with most of the
mass being in 0.4 $\mu$m grains. However, Bianchi \& Schneider (2007)
predict that amorphous carbon grains formed in supernova ejecta 
have grain sizes less than 0.1 $\mu$m and that oxide grains are 
smaller than 0.01 $\mu$m. All of these estimates are considerably
smaller than 30 $\mu$m: apparently most supernova
grains are too small to raise the injection efficiencies
significantly. 

 Bernatowicz, Croat \& Daulton (2006) find that most presolar SiC 
grains formed around carbon AGB stars fall in the size range of 
0.1 to 1 $\mu$m, with some as large as 6 $\mu$m (Amari et al. 1994). 
The carrier grains of SLRIs such as $^{26}$Al, $^{41}$Ca, $^{53}$Mn, 
and $^{60}$Fe are likely to be oxide grains though, not SiC, and only a 
relatively few oxide grains have been found to date. Current analytical 
techniques preclude the isotopic identification of presolar grains 
much smaller than $\sim$ 0.1 $\mu$m. While it thus appears that AGB
stars produce somewhat larger grains than supernova remnants,
even these grains do not appear to be large enough to raise the
injection efficiencies by a significant factor. Hence we conclude
that injection efficiencies calculated purely on the basis of
gas-phase injection (Table 1), while lower bounds, appear 
to be close enough to the correct results to rule out AGB stars
as the source of the solar system's SLRIs.

\section{Conclusions}

 A new set of models with varied shock densities and thicknesses has 
shown that injection efficiencies $f_i$ and dilution factors $D$ 
can be increased by large factors ($>$ 10 and $>$ 1000, respectively),
large enough to maintain the viability of this SLRI injection mechanism.
Observations of supernova remnants and planetary nebulae imply that
while the former shock fronts are thin enough to be suitable for 
SLRI injection, the latter are not. These results lend support to
previous studies that have favored a supernova over an AGB star for
the source of the solar system's SLRIs. Huss et al. (2009) found 
that an intermediate-mass AGB star could explain the production of
$^{26}$Al, $^{41}$Ca, $^{60}$Fe, but not that of $^{53}$Mn. 
Kastner \& Myers (1994) pointed out that AGB stars are seldom
found in the vicinity of star-forming regions, so the chances
of SLRI injection from a planetary nebula wind into a dense cloud core 
are small. The culprit appears to have been a long-forgotten supernova
remnant that swept through the galaxy $\sim$ 4.56 Gyr ago.

\acknowledgements

 This paper is dedicated to the memory of Elizabeth A. Myhill, who 
began the FLASH code effort at DTM. We thank the referee, Shogo Tachibana, 
for comments that have led to the improvment of the manuscript. 
The calculations were performed on the dc101 cluster at DTM.
This research was supported in part by NASA Origins of Solar Systems 
grant NNX09AF62G and NASA Planetary Geology \& Geophysics grant 
NNX07AP46G, and is contributed in part to NASA Astrobiology Institute
grant NCC2-1056. The software used in this work was in part
developed by the DOE-supported ASC/Alliances Center for 
Astrophysical Thermonuclear Flashes at the University of Chicago.

\clearpage
\begin{deluxetable}{lcccccccc}
\tablecaption{Injection efficiencies $f_i$ (top three rows) and
dilution factors $D$ (bottom three rows) as a function of shock 
density and thickness factors compared to the standard values 
of $3.6 \times 10^{-20}$ g cm$^{-3}$ and 0.003 pc, respectively.
For a supernova shock front slowed by snowplowing, the $D$ values
will be further reduced. Asterisks denote clouds that did not collapse.
\label{tbl-1}}
\tablewidth{0pt}
\tablehead{\colhead{shock density $\times$} 
& \colhead{} 
& \colhead{0.1} 
& \colhead{1}
& \colhead{10}
& \colhead{100}
& \colhead{200}
& \colhead{400}
& \colhead{800} }
\startdata

thickness $\times$ 0.1 & $f_i=$ &  --   & 2E-4  & 2E-3  & 1E-2  & 2E-2 & 4E-2 & 6E-2* \\

thickness $\times$  1  & $f_i=$ &  6E-5 & 1E-3  & 3E-3  & 1E-2* &  --  &  --  &  --   \\

thickness $\times$ 10  & $f_i=$ &  --   & 4E-4  & 2E-3* &  --   &  --  &  --  &  --   \\

\hline

thickness $\times$ 0.1 & $D=$ &  --   & 1E-8  & 1E-6  & 7E-4  & 3E-3 & 1E-2 & 3E-2* \\

thickness $\times$  1  & $D=$ &  4E-8 & 7E-6  & 2E-4  & 7E-3* &  --  &  --  &  --   \\

thickness $\times$  10 & $D=$ &  --   & 3E-5  & 1E-3* &  --   &  --  &  --  &  --   \\

\enddata
\end{deluxetable}

\clearpage

\begin{figure}
\vspace{-1.0in}
\plotone{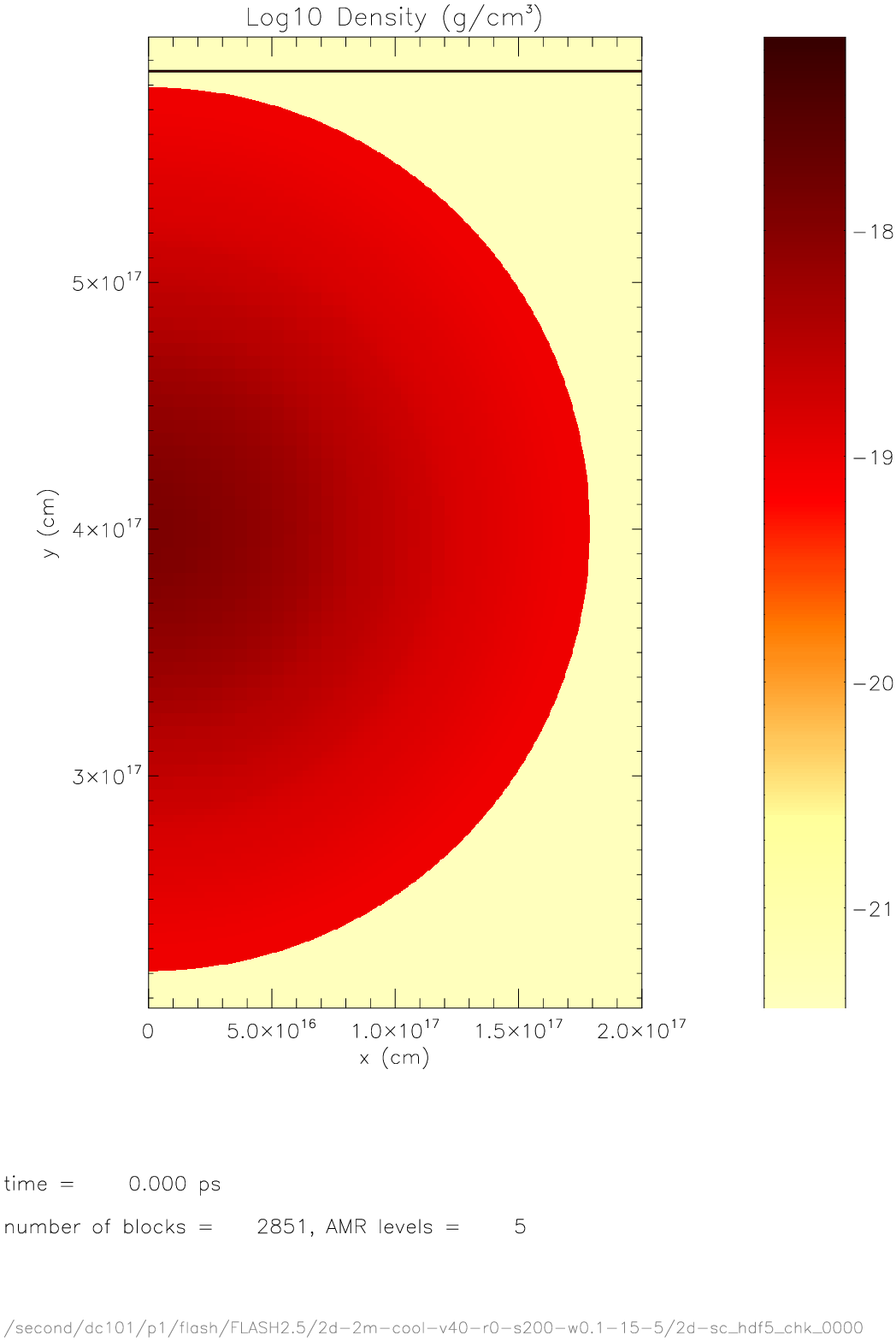}
\vspace{+0.25in}
\caption{Initial log density distibution for model 200-0.1. Black 
contours (top) enclose shock front material (representing SLRI), 
which is moving downward and is about to strike the target cloud.
The shock front has a thickness of 0.003 pc and a density 
of $3.6 \times 10^{-20}$ g cm$^{-3}$. Left side is the symmetry 
axis, with $R$ horizontal and $Z$ vertical.}
\end{figure}

\begin{figure}
\vspace{-1.0in}
\plotone{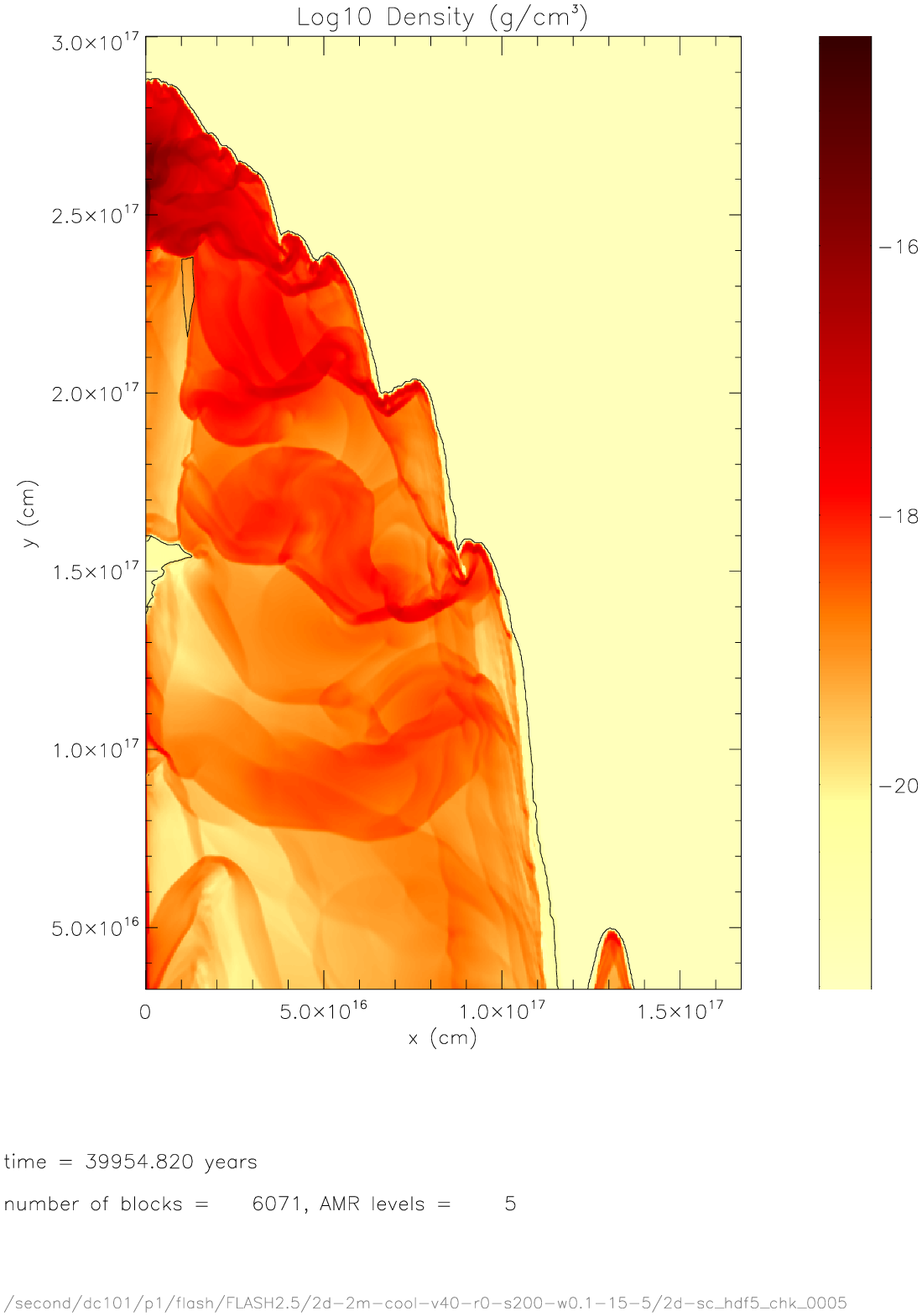}
\vspace{+0.25in}
\caption{Cloud density after 0.04 Myr, plotted in the same manner as
in Figure 1. Instabilities at the shock-cloud interface have
injected shock wave material throughout most of the target cloud 
while ablating the outer regions into the downstream flow.}
\end{figure}

\begin{figure}
\vspace{-1.0in}
\plotone{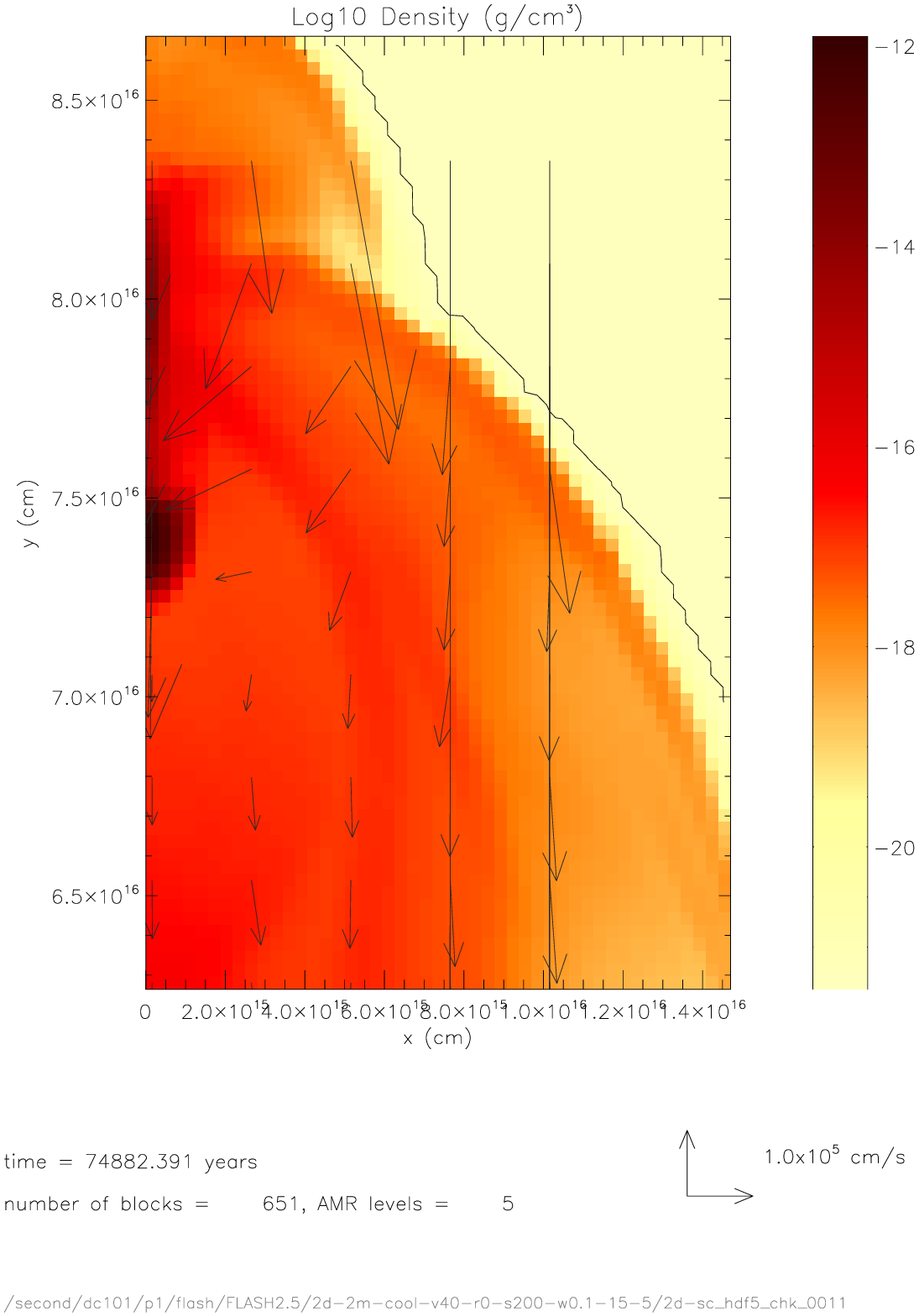}
\vspace{+0.25in}
\caption{Inner collapsing region of the cloud after 0.075 Myr, plotted
as before, showing the dense gas along the symmetry axis. Velocity
vectors are plotted for every eighth AMR grid point.}
\end{figure}

\begin{figure}
\vspace{-1.0in}
\plotone{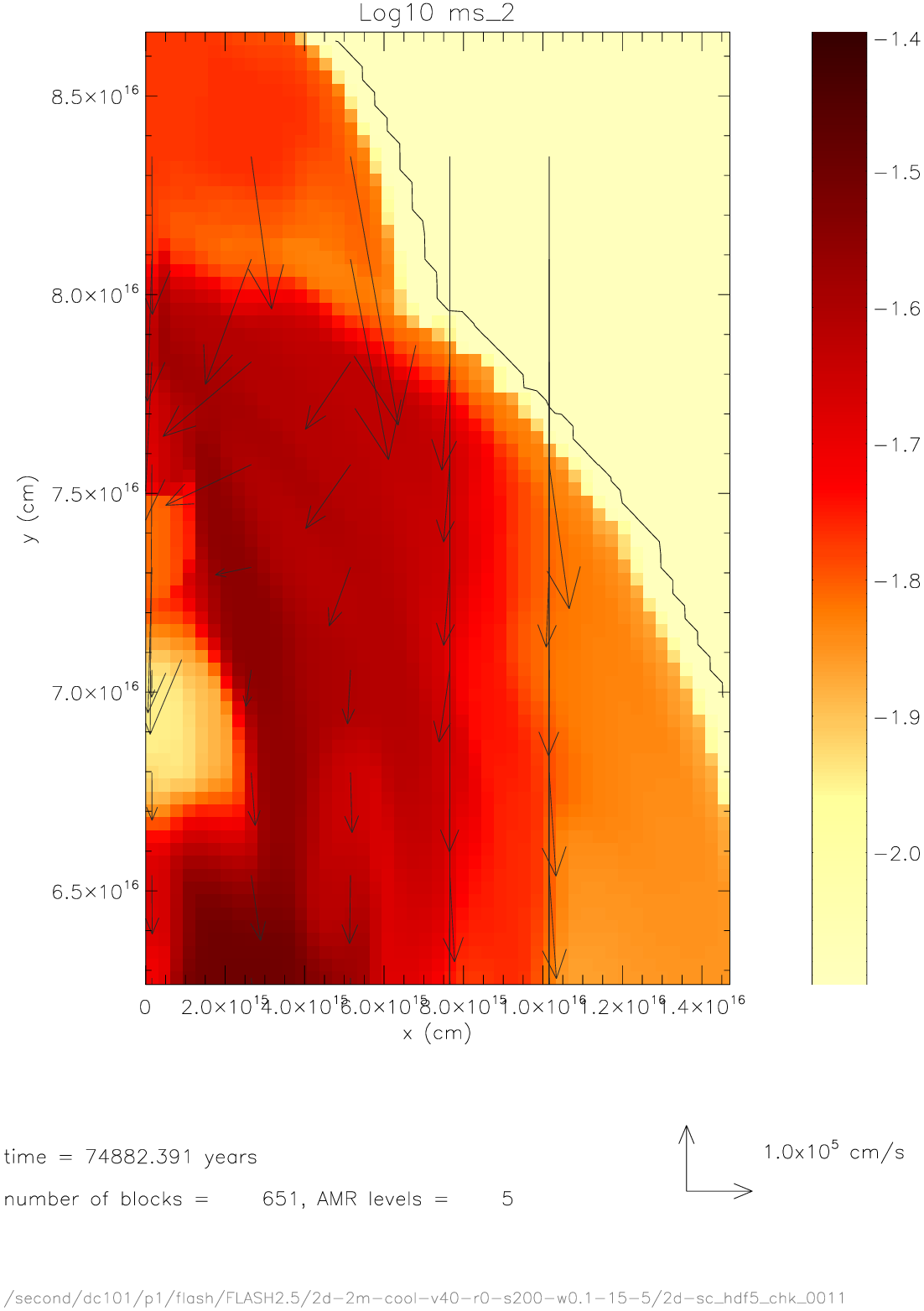}
\vspace{+0.25in}
\caption{Log shock color density (SLRIs) distribution after 0.075 Myr, 
showing the same region as in Figure 3. SLRIs have been injected inside
the growing protostar and more are infalling onto it.}
\end{figure}

\end{document}